\documentclass[twocolumn,prl,showpacs,superscriptaddress,amsmath,amssymb]{revtex4-1}
\usepackage{graphicx}% Include figure files
\usepackage{dcolumn}% Align table columns on decimal point
\usepackage{bm}% bold math

\begin{document}

\title{Response of acoustic phonons to charge and orbital order in the 50\% doped bilayer manganite LaSr$_2$Mn$_2$O$_7$}% Force line breaks with \\

\author{F. Weber}
\email{frank.weber@kit.edu}
\affiliation{Materials Science Division, Argonne National Laboratory, Argonne, Illinois, 60439, USA}
\affiliation{Karlsruher Institut f\"ur Technologie, Institut f\"ur Festk\"orperphysik, P.O.Box 3640, D-76021 Karlsruhe, Germany}
\author{S. Rosenkranz}
\author{J.-P. Castellan}
\author{R. Osborn}
\author{H. Zheng}
\author{J. F. Mitchell}
\affiliation{Materials Science Division, Argonne National Laboratory, Argonne, Illinois 60439, USA}
\author{Y. Chen}
\author{Songxue Chi}
\affiliation{NIST Center for Neutron Research, National Institute of Standards and Technology, Gaithersburg, Maryland 20899, USA}
\affiliation{Department of Materials Science and Engineering, University of Maryland, College Park, Maryland 20742, USA}
\author{J. W. Lynn}
\affiliation{NIST Center for Neutron Research, National Institute of Standards and Technology, Gaithersburg, Maryland 20899, USA}
\author{D. Reznik}
\affiliation{Karlsruher Institut f\"ur Technologie, Institut f\"ur Festk\"orperphysik, P.O.Box 3640, D-76021 Karlsruhe, Germany}
\affiliation{Department of Physics, University of Colorado at Boulder, Boulder, Colorado 80309, USA}

%\author{F. Weber}
%\email{fweber@anl.gov}
%\affiliation{%

%\author{F. Weber}
%\author{A. Kreyssig}%
%\affiliation{%
%\affiliation{

\date{\today}% It is always \today, today,

\begin{abstract}
We report an inelastic neutron scattering study of acoustic phonons in the charge and orbitally ordered bilayer manganite LaSr$_2$Mn$_2$O$_7$. For excitation energies less than $15\,\rm{meV}$, we observe an abrupt increase (decrease) of the phonon energies (linewidths) of a transverse acoustic phonon branch at $\mathbf{q} = (h, h, 0)$, $h\le0.3$, upon entering the low temperature charge and orbital ordered state ($T_{COO} = 225\,\rm{K}$). This indicates a reduced electron-phonon coupling due to a decrease of electronic states at the Fermi level leading to a partial removal of the Fermi surface below $T_{COO}$ and provides direct experimental evidence for a link between electron-phonon coupling and charge order in manganites.
\end{abstract}

% CMR, phonon measurements, EPI/lattice dynamics, Neutron Inelastic scattering
\maketitle

The complex phase diagrams of many transition metal oxides highlight the strong interplay and competition between lattice, spin and charge degrees of freedom \cite{Dagotto05,Bonn06,Schiffer95,Li07}. Among the various different ground states, the so-called CE-type \cite{Wollan55} charge and orbital order (COO) has particularly attracted scientific interest. Long range COO is the ground state of half-doped manganites \cite{Li07,Wollan55,Goodenough55}, and short-range CE-type COO is believed to play a crucial role for colossal magnetoresistance at lower doping \cite{Millis95,Millis96}. The origin of the charge modulation is typically attributed to the ordering of Mn$^{3+}$ and Mn$^{4+}$ ions\cite{Goodenough55} producing Jahn-Teller-type distortions of the oxygen octahedra around the Mn$^{3+}$ sites. Although these lattice distortions have been verified experimentally, the charge disproportionation has been argued to be much smaller than 1 \cite{Radaelli97,Argyriou00,Coey04,Herrero04}. More recently, it has been shown that the doping dependence of the ordering wavevector for pseudocubic manganites near half doping can be described in a charge-density-wave (CDW) picture \cite{Milward05} and, indeed, experimental evidence for such a scenario was found in La$_{0.5}$Ca$_{0.5}$MnO$_3$ \cite{Cox07} and Pr$_{0.48}$Ca$_{0.52}$MnO$_3$ \cite{Cox08}.

Although the importance of electron-lattice interaction for manganites and, in particular, the CE-type ordered state is based on theoretical considerations \cite{Millis95,Millis96} as well as experimental observations \cite{Manella05}, detailed experimental information on phonon dispersions and electron-phonon coupling in manganites is scarce \cite{Reichardt99,Zhang02,Reznik05,Weber09}. Here, we present results of an investigation of acoustic phonons in the CE-type COO bilayer manganite La$_{2-2x}$Sr$_{1+2x}$Mn$_2$O$_7$ ($x = 0.5$). We found that phonons are clearly sensitive to the onset of COO at $T_{COO} = 225\,\rm{K}$, but in a way not expected from the standard CDW picture \cite{Chan73,Gruener88}.

We chose the double layer manganite despite the large crystallographic unit cell because it lacks structural complications of its pseudocubic counterparts such as twinning and tilted MnO$_6$ octahedra \cite{Reznik05}. As phonon softening is often observed as a precursor to a structural phase transition at the ordering wavevector, our focus was on acoustic phonon dispersions along directions that include wavevectors where superstructure peaks were reported \cite{Argyriou00,Kubota99,Ling00}. To this end, we measured the \textit{ab}-plane polarized transverse acoustic (TA) phonon branch in the crystallographic (110) direction, which includes the COO wavevector $\mathbf{q}_{COO} = (0.25, 0.25, 0)$, and the longitudinal acoustic branch in the (100) direction. In the latter we looked for anomalous behavior associated with the presence of short-range polaron correlations with a wavevector $\mathbf{q} = (0.3, 0, 1)$ \cite{Vasiliu99,Campbell01} by comparing data from $\mathbf{Q} = (2.3,0,0)$ and $(2.3,0,1)$.

The neutron scattering experiments were performed on the BT-7 thermal triple-axis spectrometer at the NIST Center for Neutron Research. Pyrolytic graphite was used as both monochromator (vertically focused) and analyzer (horizontally focused). We used a final energy of $E_f = 14.7\,\rm{meV}$ and a graphite filter in the scattered beam in order to suppress higher order scattering. The resolution was calculated based on the Cooper-Nathans formalism \cite{Cooper67} and a force-constant model for the three-dimensional phonon dispersion developed in an earlier investigation of La$_{1.2}$Sr$_{1.8}$Mn$_2$O$_7$ \cite{Reichardt11}.

Our single crystal sample of LaSr$_2$Mn$_2$O$_7$ was melt grown in an optical image furnace \cite{Mitchell97} having the shape of a cylinder of $0.4\,\rm{cm}$ in diameter and $2\,\rm{cm}$ length. The crystal was mounted in a standard closed-cycle refrigerator allowing measurements down to $T = 5\,\rm{K}$ and up to room temperature. Measurements were carried out in the $h00-0k0$ and $h00-00l$ scattering planes. The components $(Q_h, Q_k, Q_l)$ are expressed in reciprocal lattice units (rlu) of $(2\pi/a, 2\pi/b, 2\pi/c)$ with $a = b = 3.88\,$\AA$\,$ and $c = 19.8\,$\AA. Phonon measurements were performed in the constant-\textbf{Q} mode, $\mathbf{Q} = \mathbf{\tau} + \mathbf{q}$, and $\mathbf{\tau}$ is a reciprocal lattice point.

The ground state of LaSr$_2$Mn$_2$O$_7$ has been investigated extensively by both neutron and x-ray scattering. Whereas early neutron scattering experiments \cite{Argyriou00,Kubota99,Ling00} reported COO only between $100$ and $200\,\rm{K}$, a later study\cite{Li07} showed that CE-type COO is indeed the ground state of LaSr$_2$Mn$_2$O$_7$. However, already very small deviations from $x = 0.5$ result in an A-type antiferromagnetic ground state. So far, CE-type COO at low temperatures has been observed only in tiny single crystals ($\sim 1\,\rm{mg}$)\cite{Li07} not suited for inelastic neutron scattering experiments. Therefore, we carefully characterized our sample in terms of emerging superstructures cooling from room temperature to $T = 5\,\rm{K}$.

\begin{figure}
\begin{center}
\includegraphics[width=0.95\linewidth]{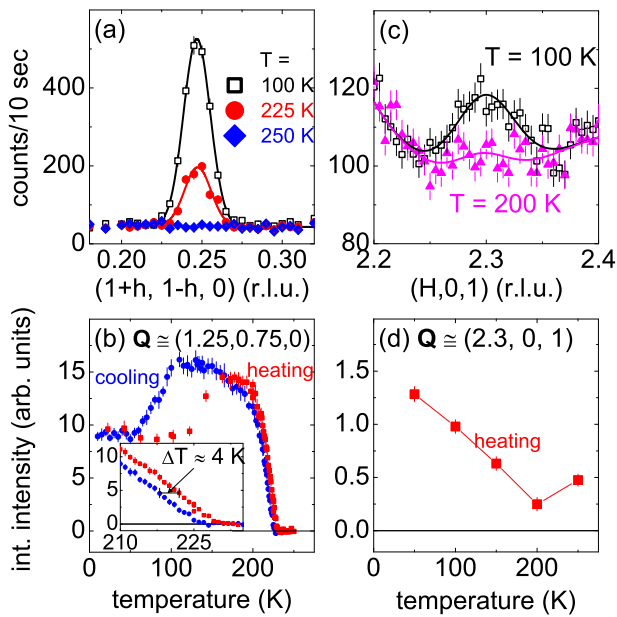}
\caption{\label{fig_1} (Color online)  Elastic \textbf{Q} scans (heating cycles) at \textit{(a)} $\mathbf{Q} = (1.25, 0.75, 0)$ and \textit{(c)} $(2.3, 0, 1)$. Solid lines are Gaussian fits for the given temperatures shown on estimated experimental backgrounds. \textit{(b)(d)} Integrated intensities of the \textbf{Q} scans around $\mathbf{Q} = (1.25, 0.75, 0)$ and $(2.3, 0, 1)$, respectively, as a function of temperature. The inset in \textit{(b)} shows the temperature hysteresis near the phase transition temperature $T_{COO}\approx 225\,\rm{K}$.}
\end{center}
\end{figure}

We show temperature dependent measurements of the COO superlattice peak at $\mathbf{Q} = (1.25, 0.75, 0)$ in Fig.~\ref{fig_1}. The onset of COO in our sample is at $T_{COO} = 225\,\rm{K}$ in agreement with previous work \cite{Li07}. On cooling below $T \approx 100\,\rm{K}$, the peak intensity decreases by about 40\%. Furthermore, we observe a large temperature hysteresis between $T = 50\,\rm{K}$ and $150\,\rm{K}$. Both facts indicate a partially reentrant behavior. Previous data \cite{Li07} on reentrant samples show a strong competition with an A-type antiferromagnetic state at low temperatures with a similarly large hysteresis in the same temperature range. More recently, it was shown that the CE-type orbital ordering as we observe it for $T\le 225\,\rm{K}$ is susceptible to both CE-type and A-type antiferromagnetism \cite{Lee11}. The reduced integrated intensity of the COO superlattice peak at low temperatures can then be attributed to ferromagnetic fluctuations in the \textit{ab} plane of an A-type antiferromagnet, which favor double exchange and, therefore, slowly melt the Mn$^{3+}$-Mn$^{4+}$ charge-order, which persists in the CE-type antiferromagnetic regions.

We also checked for the presence of short-range polaron correlations with a wavevector $\mathbf{q} \approx (0.3, 0, 1)$. In Fig.~\ref{fig_1}c we plot elastic scans taken at $\mathbf{Q} = (2+h, 0, 1)$, $h = 0.2 - 0.4$, and $T = 100$ and $200\,\rm{K}$. We clearly observe the presence of this type of superstructure in the partially reentrant state at low temperatures, but with a finite correlation length $\xi \approx 40\,$\AA$\,$ and an amplitude that is an order of magnitude smaller than the one observed for the COO peak. The temperature dependence of the polaron peak at $\mathbf{q} \approx (0.3, 0, 1)$ as observed in a heating cycle is roughly opposite to that of the COO peak at $\mathbf{q}_{COO}$, i.e. the polaron peak is barely present when the COO peak reaches its maximum intensity.

From the measurements presented in Fig.~\ref{fig_1} it is evident that CE-type COO is reduced at low temperatures. However, the measured phonons (see below) react only to the onset of COO at $T_{COO} = 225\,\rm{K}$ and show no response to the reduction of COO at lower temperatures. Thus, we believe that CE-type COO is representative of the ground state of our specimen as far as phonons are concerned. This view is corroborated by the fact that the competing A-type antiferromagnetic phase in principle supports and only slowly melts the orbital order via an increased double-exchange rate \cite{Lee11}.

\begin{figure}
\begin{center}
\includegraphics[width=0.95\linewidth]{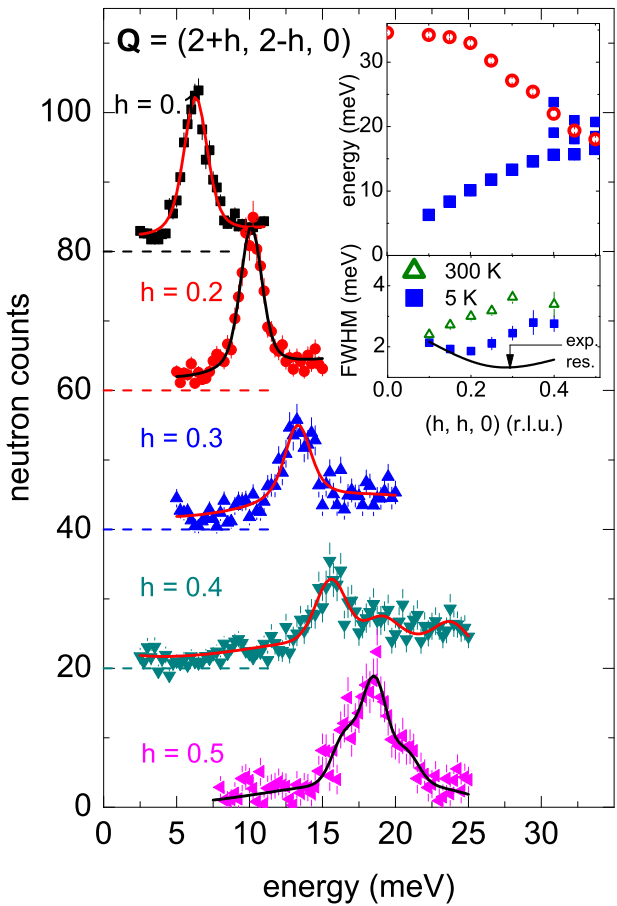}
\caption{\label{fig_2} (Color online) Constant \textbf{Q} scans of the TA phonon at $\mathbf{Q} = (2+h, 2-h, 0)$, $h = 0.1 - 0.5$, at $T = 5\,\rm{K}$. Solid lines are Lorentzian fits convoluted with the instrumental resolution on an experimental background. Dashed horizontal lines indicate the respective scan base line. The insets show energies measured at $\mathbf{Q} = (2+h, 2-h, 0)$ (squares) and $(3-h, 3+h, 0)$ (circles) \textit{(upper panel)} and phonon linewidths (FWHM) at $T=300\,\rm{K}$ (squares) and $T=5\,\rm{K}$ (triangles) of the TA phonon \textit{(lower panel)}. The black line is the calculated resolution.}
\end{center}
\end{figure}

We investigated the acoustic and low lying optic phonons along the transverse (110) direction, i.e. $\mathbf{q} = (+h, -h, 0)$, where $h=0.25$ corresponds to the symmetry and wavevector positions of the COO superlattice peak. Raw data of constant \textbf{Q} scans at $T = 5\,\rm{K}$ for the TA phonon mode at $\mathbf{Q} = (2+h, 2-h, 0)$ are shown in Fig.~\ref{fig_2}. For $h \le 0.3$, a single well-defined excitation is observed. For larger wave vectors additional phonon peaks start to develop and finally three peaks can be distinguished at the zone boundary, i.e. $h = 0.5$. The fitted energies agree well with previous measurements on La$_{1.2}$Sr$_{1.8}$Mn$_2$O$_7$ \cite{Reichardt11}. Near the zone boundary, our measurements show additional peaks due to optic phonon branches, which come close in energy to the acoustic dispersion. This is corroborated by measurements in a different Brillouin zone adjacent to the zone center wavevector $\tau = (3, 3, 0)$, where the structure factors of the optic phonons are much stronger and acoustic phonons have practically zero intensity (see inset in Fig.~\ref{fig_2}). In the second inset in Fig.~\ref{fig_2} we show the observed linewidths (full width at half maximum) of the TA phonon compared to the calculated resolution for our experimental setup. It is evident that there is a general reduction of the linewidths at $T<T_{COO}$. However, modes at $T=5\,\rm{K}$ with $\mathbf{q} \ge (0.25, 0.25, 0)$ still have a significant intrinsic linewidth.

\begin{figure}
\includegraphics[width=0.95\linewidth]{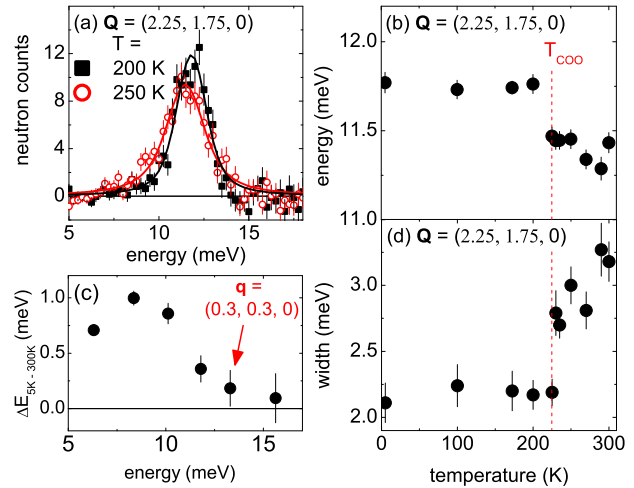}
\caption{\label{fig_3} (Color online) Temperature dependent \textit{(a)} background subtracted data, \textit{(b)} energy and \textit{(d)} Lorentzian linewidth   (full-width at half maximum) of the transverse acoustic phonon at $\mathbf{Q} = (2+h, 2-h, 0)$, $h = 0.25$. Energies and linewidths were extracted from Lorentzian fits (solid lines) to the data convoluted with a Gaussian energy resolution ($\approx 2\,\rm{meV}$). The dashed vertical lines in \textit{(b)} and \textit{(d)} mark the onset of orbital order at $T_{COO} = 225\,\rm{K}$. \textit{(c)} Energy shift of TA phonons at $\mathbf{Q} = (2+h, 2-h, 0)$, $0.1 \le h \le 0.35$, between $T = 5$ and $300\,\rm{K}$ plotted versus phonon energy at $T = 5\,\rm{K}$.}
\end{figure}

We made measurements at various temperatures $5\,\rm{K} \le T \le 300\,\rm{K}$ and $\mathbf{Q} = (2+h, 2-h, 0)$. Figure \ref{fig_3}a shows the background subtracted data at the ordering wave vector of the COO, $h = 0.25$, at $\Delta T = 25\,\rm{K}$ below and above the transition temperature $T_{COO} = 225\,\rm{K}$. The phonon energy increases and the linewidth decreases on entering the COO state. As a function of temperature, both the energy and linewidth at $h = 0.25$ show sudden jumps at $T = T_{COO}$ (Figs.~\ref{fig_3}b,d). Figure \ref{fig_3}c shows that the effect increases on an absolute scale for phonons with decreasing energies, i.e. decreasing $h$. On the other side, the energy shift between low and high temperatures vanishes (within experimental error) for a phonon energy of roughly $15\,\rm{meV}$ or $h > 0.3$. We note that no temperature dependence was detected for the investigated optic branch between $T = 5$ and $300\,\rm{K}$, which further indicates that the observed effect is restricted to low energy acoustic phonons.

Measurements of the longitudinal acoustic phonon at wavevectors corresponding to the wavevector of the short-range superstructure at $\mathbf{q} = (0.3, 0, 1)$ agreed well with shell model calculations developed for the compound with $x=0.4$ that have been discussed elsewhere\cite{Weber09}. We did not observe any particular temperature or wavevector dependence at this \textbf{q} position. We note, however, that the statistical uncertainty here is larger than for the TA phonon data and changes of less than $0.2\,\rm{meV}$ in the phonon energy could not determined.

Summarizing our experimental results, we observe a jump in the TA phonon energy and linewidth at $T_{COO}$. The effect is not localized at the ordering wavevector, i.e. $\mathbf{q}_{COO} = (0.25, 0.25, 0)$, but occurs over an extended range in momentum transfer, as long as the energy is lower than $15\,\rm{meV}$ or $\mathbf{q} \le (0.3, 0.3, 0)$.

The phonon linewidth can be roughly divided into an anharmonic part and the electronic contribution to the phonon linewidth, i.e. electron-phonon coupling (EPC). Anharmonic interactions typically result in a gradual increase of the phonon linewidth with increasing temperature. In principle, this can happen in a more abrupt manner, when a rigid order suddenly relaxes into an unordered state. However, this effect should not be limited to a certain range of phonon wavevectors or energies. Therefore, the observed abrupt changes in the phonon linewidths and energies in LaSr$_2$Mn$_2$O$_7$ cannot be explained by a sudden increase of anharmonic contributions.

EPC for a particular phonon mode requires the existence of electronic states close to the Fermi energy $E_F$, which can be excited by the phonon to unoccupied states above $E_F$. If these decay channels are frozen out, the phonon lifetime increases and the linewidth is reduced. This is a well known effect in, e.g., conventional superconductors for phonon energies below the superconducting gap value $2\Delta_{SC}$ \cite{Axe73,Shapiro75,Weber10}. Further, the participating electronic states have to be connected by the phonon wavevector \textbf{q}. Unfortunately, information about the Fermi surface in the half-doped bilayer manganite is scarce. Angle-resolved photoemission spectroscopy reported only experiments in the COO state \cite{Sun11}. Here, the Fermi surface is such that phonon vectors with $\mathbf{q} < (0.25, 0.25, 0)$ cannot connect different states at $E_F$. On the other hand, calculations of the electronic band structure via density-functional theory \cite{Saniz08} in not-charge-ordered LaSr$_2$Mn$_2$O$_7$ show the presence of a small electron pocket around the center of the Brillouin zone ($\Gamma$ point) in addition to the Fermi surface observed in the low temperature phase by angle-resolved photoemission spectroscopy \cite{Sun11} allowing Fermi surface spanning wave vectors with $\mathbf{q} < (0.25, 0.25, 0)$. Hence, the removal of this pocket across the COO phase transition might lead to a sudden loss of electronic decay channels for $\mathbf{q} < (0.25, 0.25, 0)$ and explain the observed effect.
The fact that the observed effects become much weaker at $\mathbf{q} = (0.25, 0.25, 0)$ and are not detectable anymore at $\mathbf{q} \ge (0.35, 0.35, 0)$ can be understood in terms of the published angle-resolved photoemission spectroscopy data \cite{Sun11}: Wave vectors $\mathbf{q} \ge (0.25, 0.25, 0)$ can still connect parts of the Fermi surface in the COO state. Thus, phonons with these wavevectors lose only part of their electronic decay channels. This is corroborated by our results for the linewidths of the TA phonon modes (Fig.~\ref{fig_2}): Despite the sudden decrease of the linewidth of the TA phonon at $\mathbf{q} = (0.25, 0.25, 0)$ (Fig.~\ref{fig_3}), it keeps a significant intrinsic linewidth even at the lowest temperatures. Thus, our results demonstrate the presence of significant EPC and a direct response of EPC to the COO phase transition in manganites.

We note that in the recent literature the CE-type ordered low temperature state of half-doped La-Ca manganites is discussed in terms of a charge-density wave \cite{Coey04,Cox07,Herrero04,Calvani98,Nucara08}. In this scenario, a gap $\Delta$ in the electronic excitation spectrum opens at the phase transition \cite{Calvani98}. Such a gap opening could cause a reduced phonon linewidth and changes in the energy, if the latter are smaller than $2\Delta$ \cite{Shapiro75}. Our results would be consistent with $2\Delta \approx 15\,\rm{meV}$ at $T = 5\,\rm{K}$. So far, however, there is no microscopic evidence from electronic probes for the existence of such an energy gap in LaSr$_2$Mn$_2$O$_7$ \cite{Dessau11}. Furthermore, CDW compounds typically exhibit phonon softening at the ordering wavevector as predicted by standard weak-coupling theory. Some known CDW compounds do not follow this phenomenology exactly. For example, in NbSe$_2$, phonons soften over a relatively large range of wave vectors around the ordering wavevector \cite{Weber11}: NbSe$_3$, which is believed to be in the strong-coupling regime, shows not softening but line broadening at the CDW wave vector on approach to the transition \cite{Requardt02}. But all CDW compounds exhibit some anomalous phonon behavior tied to the ordering wavevector. In contrast, acoustic phonons in LaSr$_2$Mn$_2$O$_7$ do not show a Kohn anomaly nor an enhanced linewidth at the ordering wavevector, which rules out the conventional CDW picture.

In conclusion, we investigated acoustic phonons in the presence of long- and short-range charge correlations in CE-type COO ordered LaSr$_2$Mn$_2$O$_7$ via inelastic neutron scattering. We found a clear response to the onset of COO at $T_{COO} = 225\,\rm{K}$ in the TA branch in the (110) direction. For this branch, the phonon linewidths are significantly reduced for phonon modes with excitation energies smaller than $15\,\rm{meV}$ $[\mathbf{q} \le (0.3, 0.3, 0)]$. This is direct evidence for a link between charge order and electron-phonon coupling in this compound. The observed wavevector dependence of the effect at the phase transition is not consistent with a conventional CDW mechanism of the charge ordering.

%Our observations can be explained by a reduction of electronic states at the Fermi level consistent with the reported jump in resistivity at $T_{COO}$ \cite{Li07}. That only phonon modes at small wavevectors are affected indicates a removal of a certain part of the Fermi surface on entering the COO state. We also discussed the possibility of a gap $2\Delta(T=5\,\rm{K}) \approx 15\,\rm{meV}$  with respect to observations in pseudo-cubic La$_{0.5}$Ca$_{0.5}$MnO$_3$ \cite{Calvani98}. Detailed microscopic models based on the full electronic structure are necessary to better understand our results.

\begin{acknowledgments}
Work at Argonne was supported by U.S. Department of Energy, Office of Science, Office of Basic Energy Sciences, under Contract No. DE-AC02-06CH11357.
\end{acknowledgments}

%\bibliographystyle{apsrev.bst}
%\bibliography{bibtex_endnote}

\end{document}